\documentclass[aps,prl,notitlepage,twocolumn,10pt, superscriptaddress]{revtex4-1}

\usepackage{titlesec}

\usepackage{graphicx}
\usepackage{array}
\usepackage{float}
\usepackage{amsmath}
\usepackage{amssymb}
\usepackage{verbatim}
\usepackage{amsmath}
\usepackage{physics}
\usepackage{manfnt}
\usepackage{titlesec}
\usepackage[colorlinks=true]{hyperref}
\usepackage{siunitx}

\begin{document}
	
\title{Spin Seebeck Effect near the Antiferromagnetic Spin-Flop Transition}
\author{Derek Reitz}
\affiliation{Department of Physics and Astronomy, University of California, Los Angeles, California 90095, USA}
\author{Junxue Li}
\author{Wei Yuan}
\author{Jing Shi}
\affiliation{Department of Physics and Astronomy, University of California, Riverside, California 92521, USA}
\author{Yaroslav Tserkovnyak} 
\affiliation{Department of Physics and Astronomy, University of California, Los Angeles, California 90095, USA}
\begin{abstract}	
We develop a low-temperature, long-wavelength theory for the interfacial spin Seebeck effect (SSE) in easy-axis antiferromagnets. The field-induced spin-flop (SF) transition of N\'eel order is associated with a qualitative change in SSE behavior: Below SF, there are two spin carriers with opposite magnetic moments, with the carriers polarized {\em along the field} forming a majority magnon band. Above SF, the low-energy, ferromagnetic-like mode has magnetic moment {\em opposite the field.} This results in a sign change of the SSE across SF, which agrees with recent measurements on Cr$_2$O$_3$/Pt and Cr$_2$O$_3$/Ta devices [Li \textit{et al.,} \textit{Nature} \textbf{578,} 70 (2020)]. In our theory, SSE is due to a N\'eel spin current below SF and a magnetic spin current above SF. Using the ratio of the associated N\'eel to magnetic spin-mixing conductances as a single constant fitting parameter, we reproduce the field dependence of the experimental data and partially the temperature dependence of the relative SSE jump across SF.
\end{abstract}

\maketitle

\textit{Introduction.}|SSE involves transfer of spin angular momentum between a magnet and a metal via thermal spin fluctuations at their interface. In a typical experiment, a heat flux injected across the interface pumps a spin current into the metal, which is then converted into a transverse electric voltage $V_{\mathrm{SSE}}$ by spin-orbit interactions. This spin-current generation can be broadly attributed to two sources: One is due to a thermal gradient inside the magnet, which produces bulk magnon transport \cite{Adachi_2010, rezende2014magnon, rezende2016bulk, Flebus_2017, Prakash_2018, luo2019} and results in interfacial spin accumulation. The other is due to the interfacial temperature discontinuity, which produces spin pumping directly \cite{xiao2010}.

SSE has been studied in ferromagnets \cite{slachter2010, uchida2010}, ferrimagnets \cite{miao_2016, geprags2016, ohnuma_2013}, paramagnets \cite{wu2015paramagnetic, Li_2019, yamamoto2019}, and recently in antiferromagnets \cite{Seki_2015, Wu_2016, li2020, Rezende_2016, troncoso2019} as well as noncollinear magnets \cite{flebus_2019_noncollinear, ma2019longitudinal}. The sign of $V_{\mathrm{SSE}}$ is determined by the polarization of the spin current along the applied magnetic field and the effective spin Hall angle of the metal detector. Fixing the spin Hall angle and the gyromagnetic ratio, the observed sign of the underlying spin current turns out to contain valuable information about the nature of spin order in the magnet and its nonequilibrium transport properties.

Collinear ferromagnets (FMs) or noncollinear systems with weak ferromagnetic order have their net spin ordering along the magnetic field, whereas the elementary low-energy magnon excitations yield average spin polarization in the opposite direction. We can also imagine another class of systems, whose intrinsic excitations form spin-degenerate bands, with the degeneracy lifted by Zeeman splitting. The majority species, polarized along the field, may then determine the sign of the spin current, thus ending up opposite to the FM case. In our formalism, uniaxial AFs fall in this latter, majority-species scenario below SF, switching to the ferromagnetic-like SSE behavior above SF. Unlike argued in Ref.~\cite{hirobe2017}, therefore, the SSE with the sign opposite to the FM case is a not a unique signature of correlated spin liquids, but can be expected to be a rather generic low-temperature signature of materials lacking FM order.

Theoretically, there is at present no consensus on the ``correct" sign of the SSE in antiferromagnets. Rezende et al. \cite{Rezende_2016} developed a magnon transport theory for uniaxial AFs below SF and concluded it falls into the majority-species scenario (i.e., SSE opposite to the FM case), but did not consider the sign when comparing their theory to experiment. Yamamoto et al. \cite{yamamoto2019} used the fluctuation-dissipation theorem in a Landau-Ginzburg theory for easy-axis AFs below SF to study SSE around the N\'eel temperature $T_N$, concluding paramagnets and AFs below SF both have the same sign, but that it is the same as FMs. Here, we determine the sign within a low-temperature, long-wavelength theory for the interfacial SSE and show it changes across SF, in agreement with recent experiments. The quantitative aspects of the SSE over a broad range of temperatures and magnetic fields also appear in general agreement with the data.

\begin{figure}
	\includegraphics[width=2.8in]{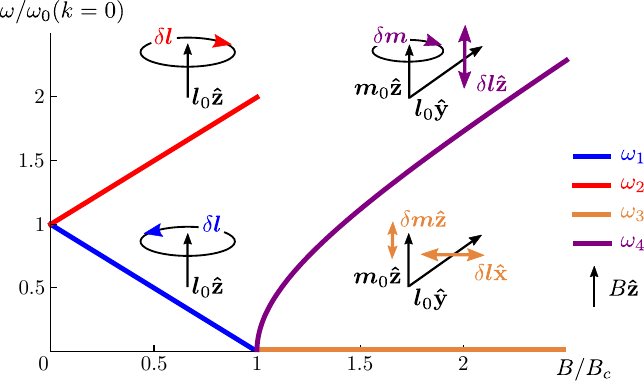}
	\caption{$k = 0$ resonance frequencies are plotted for an easy-axis AF: $\omega_1$ and $\omega_{2}$ below spin flop and $\omega_{3}$ and $\omega_{4}$ above spin flop. $B$ is the applied magnetic field, $B_c=(\gamma s)^{-1}\sqrt{K_1/\chi}$ is the spin-flop field (which is about 6~Tesla for Cr$_2$O$_3$) according to the energy~\eqref{energy}, and $\omega_{0} = \gamma B_c$ is the gap in I. The $\omega_1$ mode is right-hand circularly polarized and $\omega_{2}$ is left-hand circularly polarized in $\delta\boldsymbol{l}$ and $\delta\boldsymbol{m}$ (however the magnitude of $\delta\boldsymbol{m}$ is a factor $\chi K_1$ smaller than $\delta\boldsymbol{l}$ below SF, so it is omitted from the Figure). $\omega_{3}$ is linearly polarized in $\delta\boldsymbol{l}$ and $\delta\boldsymbol{m}$ so it does not produce spin currents \cite{beyond_nonlinear_sigma}. $\omega_{4}$ is linearly polarized in $\delta\boldsymbol{l}$ and elliptically polarized in $\delta\boldsymbol{m}$.}
	\label{fig:fig_1}
\end{figure}

\textit{Spin pumping near SF transition.}|In easy-axis AFs, when the Zeeman energy due to an applied field along the easy axis exceeds the anisotropy energy, there is a metamagnetic phase transition called spin flop (SF). Below SF (state I), the N\'eel order aligns with the easy axis, and there is a small net magnetization due to remnant longitudinal magnetic susceptibility \cite{Nordblad_1979, Foner_1963}. Dynamically, there are two circularly-polarized spin-wave modes with opposite handedness. When quantized, they correspond to magnons with magnetic moment parallel or antiparallel to the order parameter, each forming a gas (with equal and opposite chemical potentials, if driven slightly out of equilibrium \cite{flebus2019chemical}). Above SF (state II), the N\'eel order reorients into the hard plane, and the spins cant giving net magnetization along the easy axis, due to a sizeable transverse magnetic susceptibility. There are now two distinct spin-wave modes at long wavelengths: a ferromagnetic-like mode ($\omega \rightarrow \gamma B$ when applied field $B \rightarrow \infty$) and a low-energy Goldstone mode associated with the U(1)-symmetry breaking N\'eel orientation in the hard plane. See Fig.~\ref{fig:fig_1}.

The spin-current density pumped across the interface consist of the N\'eel, $\boldsymbol{J}_l$, and magnetic, $\boldsymbol{J}_m$, contributions:
\begin{equation}
\label{g_def}
\boldsymbol{J}_l = (\hbar g^{\uparrow \downarrow}_l/4\pi) \,\boldsymbol{l} \times \partial_t \boldsymbol{l},~~~\boldsymbol{J}_m = (\hbar g^{\uparrow \downarrow}_m/4\pi) \,\boldsymbol{m} \times \partial_t\boldsymbol{m},
\end{equation}
where $g^{\uparrow \downarrow}$ is the respective (real part of the dimensionless) interfacial spin-mixing conductance per unit area. Thermal agitations in the metal held at temperature $T_e$ and in the AF at $T_a$ produce contributions $\boldsymbol{J}_e$ and $\boldsymbol{J}_a$ to the spin current, respectively. The spin Seebeck coefficient $S$ can be defined as the net spin current $J_s$ (projected onto the direction of the applied field) across the interface, divided by the temperature drop $\delta T=T_a-T_e$:
\begin{equation}
S \equiv J_s/\delta T= [J_a(T_a) - J_e(T_e)]/\delta T \to \partial_TJ_a(T),
\label{s_def}
\end{equation}
in linear response, where $J_a=J_l+J_m$ and $J_e(T) = J_a(T)$, in thermal equilibrium.

In this paper, we investigate the signatures of SF in the SSE. In state I, there are two components of the N\'eel spin current that contribute oppositely to the SSE. With respect to increasing field, the (anti)parallel mode (decreases) increases in frequency. The antiparallel mode thus has greater thermal occupation at finite field, producing a net N\'eel spin current antiparallel to the field \cite{ohnuma_2013, Rezende_2016}. In state II, there is only a magnetic spin current parallel to the field from the FM-like mode. Therefore, the SSE changes sign across SF.

\textit{Spin-wave modes.}|Following standard procedure \cite{Andreev_1980}, we construct the low-energy long-wavelength theory for AF dynamics in terms of the Lagrangian density $\mathcal{L}(\boldsymbol{l}, \boldsymbol{m}) = s\boldsymbol{m}\cdot (\boldsymbol{l} \times \partial \boldsymbol{l}/\partial t) - E$. The energy density is given here by
\begin{equation}
\label{energy}
E(\boldsymbol{l}, \boldsymbol{m}) = A(\grad{\boldsymbol{l}})^2/2 + \boldsymbol{m}^2/2 \chi - K_1l_z ^2/2 - b\,\boldsymbol{m} \cdot \hat{\textbf{z}},
\end{equation}
for a bipartite easy-axis AF subjected to a collinear magnetic field. The AF state is parametrized by directional N\'eel order $\boldsymbol{l}$ and normalized spin density $\boldsymbol{m}=\mathbf{s}/s$ ($\mathbf{s}$ being the spin density and $s\equiv\hbar S/V$, for spin $S$ and volume $V$ per site), in a nonlinear $\sigma$ model with constraint $\boldsymbol{l}^2 = 1$ and $\boldsymbol{l}\cdot\boldsymbol{m} = 0$. We work well below the ordering temperature $T_N$, retaining the lowest-order gradient term of the N\'eel order with spin stiffness $A$. $\chi$ is the transverse magnetic susceptibility, $K_1$ the easy-axis anisotropy, and $b\equiv\gamma s B$, in terms of the magnetic field $B$ applied along the easy axis in the $\hat{\textbf{z}}$ direction (where $\gamma$ is the gyromagnetic ratio, whose sign is lumped into the value of $B$; \textit{i.e.}~when $\gamma < 0$, our $B$ has opposite sign to the applied field). The Euler-Lagrange equations of motion may be extended to include dissipative forces $\partial \mathcal{F} / \partial \dot{\boldsymbol{m}}$ and $\partial \mathcal{F} / \partial \dot{\boldsymbol{l}}$ from the Rayleigh dissipation functional $\mathcal{F} = \alpha \dot{\boldsymbol{l}}^2/2 + \widetilde{\alpha} \dot{\boldsymbol{m}}^2/2$, parametrized by Gilbert damping constants $\alpha$ and $\widetilde{\alpha}$.

The ground states I and II are $(\boldsymbol{l}_0,\boldsymbol{m}_0)_\mathrm{I} = (\hat{\textbf{z}},0)$ and $(\boldsymbol{l}_0,\boldsymbol{m}_0)_\mathrm{II} = (\hat{\textbf{y}},\chi b\hat{\textbf{z}})$, with the critical field $B_c$ marking the jump from I to II. Spin waves are linear excitations, $\boldsymbol{l} = \boldsymbol{l}_0 + \delta\boldsymbol{l}$ and $\boldsymbol{m} = \boldsymbol{m}_0 + \delta\boldsymbol{m}$, satisfying the equations of motion. The dispersions are 
\begin{subequations}
	\label{disp}
	\begin{eqnarray}
	&& \omega_{1k}, \omega_{2k} = \mp \gamma B + \sqrt{(\gamma B_c)^2 + (ck)^2}, \\
	&& \omega_{3k} = ck,~~~\omega_{4k} = \sqrt{\gamma^2B^2 - \gamma^2 B_c^2 + (ck)^2},
	\end{eqnarray}
\end{subequations}
where $c=s^{-1}\sqrt{A/\chi}$ is the speed of the large-$k$ AF spin waves.

The six Cartesian components of $\delta\boldsymbol{l}$ and $\delta\boldsymbol{m}$ reduce to four independent and two slave variables, after applying the nonlinear constraints. Correspondingly, there are four spin-wave modes with momentum $k$, as shown in Fig.~\ref{fig:fig_1} (for consistency of the gradient expansion, we require $k \ll a^{-1}$, the inverse lattice spacing). $\omega_{1k}$ and $\omega_{2k}$ are waves with circularly precessing $\delta\boldsymbol{l}$ and $\delta\boldsymbol{m}$ in the plane perpendicular to $\boldsymbol{l}_{0, \mathrm{I}}$. $\omega_{3k}$ has linearly polarized $\delta\boldsymbol{l}(t) \propto e^{i\omega_{3k}t} \hat{\textbf{x}}$ and $\delta\boldsymbol{m}(t) \propto (\omega_{3k} / \omega_x)  e^{i(\omega_{3k}t - \pi / 2)}\hat{\textbf{z}}$ \cite{beyond_nonlinear_sigma}. $\omega_{4k}$ has linearly polarized $\delta\boldsymbol{l}(t) \propto  e^{i\omega_{4k}t} \hat{\textbf{z}}$ and elliptically polarized $\delta\boldsymbol{m}(t) \propto (\omega_{4k}/\omega_x) e^{i\omega_{4k}t} \hat{\textbf{x}} -  \chi b e^{i(\omega_{3k}t - \pi / 2)}\hat{\textbf{y}}$, where $\omega_x\equiv 1/\chi s$. Additional anisotropy energy $-K_2l_y^2/2$ within the easy plane will slightly shift the ground states, gap $\omega_3$, and introduce ellipticities in precession. When $k_BT \gg (\hbar/s)\sqrt{K_2/\chi}$, however, these modifications are negligible \cite{k2_note}. \par

\textit{Main results.}|A thermal heat flux driven across the AF interface with a metal is given in the bulk by $-\sigma \grad{T}$ and at the interface by $-\kappa \delta T$, where $\sigma$ and $\kappa$ are, respectively, the bulk and interfacial (Kapitza) thermal conductivities. $\delta T$ here is the temperature difference between phonons in the AF and electrons in the metal, $\delta T = T_p - T_e$ \cite{xiao2010, adachi_2011}. The Kapitza resistance ($\kappa^{-1}$) is large when there is poor phonon-phonon and phonon-electron interfacial coupling. For a fixed heat flux, this results in a larger $\delta T$, which drives the local SSE. The temperature gradient $\grad{T}$ inside the magnet, furthermore, generates a bulk spin current, which flows towards the interface and contributes to the measured SSE \cite{hoffman_2013}. We will specialize to the limit, in which the local spin pumping $\propto\delta T$ dominates, which corresponds to the case of an opaque interface and/or short spin-diffusion length in the AF.

Equipped with the theory for AF dynamics, based on the Hamiltonian \eqref{energy}, we can use thermodynamic fluctuation-dissipation relations in order to convert magnetic response into thermal noise. The spin Seebeck coefficient \eqref{s_def} can then be evaluated by averaging Eqs.~\eqref{g_def} over thermal fluctuations, whose spectral features follow the spin-wave dispersions discussed above. Carrying out this program, we arrive at the following final results (with the details of the derivations discussed later): Below spin flop (state I),
\begin{equation}
\label{S_I}
S_{\mathrm{I}} = \frac{ g^{\uparrow \downarrow}_l \hbar^2}{2\pi \chi s^2}  \int \frac{ d^3 k}{(2\pi)^3} \frac{\omega_{2k}\partial_Tn_{\rm BE}(\omega_{2k})-\omega_{1k} \partial_Tn_{\rm BE}(\omega_{1k})}{\omega_{1k}+\omega_{2k}},
\end{equation}
and above spin flop (state II),
\begin{equation}
\label{S_II}
S_{\mathrm{II}} = \frac{g^{\uparrow \downarrow}_m\hbar^2 \chi\gamma B }{2\pi} \int \frac{ d^3 k}{(2\pi)^3} \omega_{4k} \partial_Tn_{\rm BE}(\omega_{4k}),
\end{equation}
where $n_{\rm BE}(\omega)=(e^{\hbar\omega/k_BT}-1)^{-1}$ is the Bose-Einstein distribution function.

We may evaluate the Seebeck coefficients analytically when $k_BT\gg\hbar\gamma B_c$. Since they are both linear in $B$, we compare the field slopes which go as $\partial_B S_{\mathrm{I}} \propto g^{\uparrow \downarrow}_l T$ and $\partial_B S_{\mathrm{II}} \propto g^{\uparrow \downarrow}_m T^3 $:
\begin{equation}
\label{v_def}
v(T) \equiv - \frac{\partial_B S_{\mathrm{I}}}{\partial_B S_{\mathrm{II}}} \approx \frac{g^{\uparrow \downarrow}_l}{g^{\uparrow \downarrow}_m} \left(\frac{\hbar/\chi s}{k_{B} T }\right)^2  \sim \frac{g^{\uparrow \downarrow}_l}{g^{\uparrow \downarrow}_m} \left(\frac{T_N}{T}\right)^2.
\end{equation}
The ratio $v(T)$ contains the square of exchange ($\propto T_N$) to thermal energy in $v(T)$ (for the complete expressions, see \cite{s_I/II}). Note that for the applicability of our long-wavelength description, we require that $T \ll T_N$, throughout.

\textit{Comparison to experiment.}|In a conventional measurement scheme, the (longitudinal) SSE is revealed in a Nernst geometry as a lateral voltage induced perpendicular to the magnetic field applied in the plane of the magnetic interface \cite{uchida2010}. This voltage is understood to arise from the inverse spin Hall effect associated with the thermally injected spin current. Normalizing the SSE voltage by the input thermal power $P_{\mathrm{in}}$, this gives
\begin{equation}
\label{VSSE}
\frac{V_{\mathrm{SSE}}}{P_{\mathrm{in}}} = S(B,T)\frac{2e}{\hbar}\frac{\lambda^*}{w t}\frac{\rho(T)}{\kappa^*(T)},
\end{equation}
where the materials-dependent interfacial spin-to-charge conversion lengthscale $\lambda^*$ can be loosely broken down into a product of an effective spin-diffusion length (a.k.a. spin-memory loss) $\lambda_{\mathrm{sd}}$ in the (heavy) normal metal and the effective spin Hall angle $\theta_{\mathrm{sH}}$, which converts the spin-current density $J_{\mathrm{s}}$ injected into the normal metal into the lateral charge-current density $J_c =  (2e/\hbar)\theta_{\mathrm{sH}} J_s$. The total charge current is $I_c = w \lambda_{\mathrm{sd}}J_c$ when $\lambda_{\mathrm{sd}} \ll t$, the thickness of the metal film, where $w$ is the heterostructure width transverse to the injected charge current. In the open circuit, the underlying spin Hall motive force \cite{uchida_2010} is balanced by the detectable voltage $V_{\mathrm{SSE}} = \rho l I_c/ w t$, along the length $l$, where $\rho$ is the normal-metal resistivity. Putting everything together and expressing the spin current in terms of the Seebeck coefficient \eqref{s_def}, we get the SSE voltage \eqref{VSSE} normalized by the input power $P_{\mathrm{in}} = \kappa (T_p - T_{\mathrm{e}})l w$. $\kappa^*=\kappa(T_p-T_e)/(T_a-T_e)$ is an effective Kapitza conductance, which can be reduced relative to $\kappa$, if the lengthscale for the magnon-phonon equilibration that controls the temperature mismatch $T_a-T_p$ in the AF is long compared to $\sigma/\kappa$.

Kapitza conductances for metal-insulator interfaces have been investigated in Refs.~\cite{Stoner_1992, stevens2005measurement, hohensee2015thermal, lu2016interfacial}, yielding nontrivial temperature dependences. The parameters for Cr$_2$O$_3$ are: $\sqrt{A}/a = (\chi \gamma s)^{-1} \approx  500$~T, $B_{\mathrm{c}} \approx 6$~T, $\gamma \approx \gamma_e$ \cite{li2020} (where $\gamma_e$ is the free-electron value), $K_2 \approx 0$ \cite{Foner_1963}; for the Cr$_2$O$_3$/Pt and Cr$_2$O$_3$/Ta devices: $w = 0.2$~mm, $t = 5$~nm, the resistivity of the strips are $\rho_{\mathrm{Pt}} \approx 7 \times 10^{-6}~\Omega\cdot$m and $\rho_{\mathrm{Ta}} \approx 9 \times 10^{-5}~\Omega\cdot$m \cite{li2020} at $T = 75$~K, we take $\lambda^*$ from spin-pumping experiments: $\lambda_{\mathrm{Pt}}^*\sim 0.1$~nm \cite{sinova_2015} and $\lambda_{\mathrm{Ta}}^* \sim -0.04$~nm \cite{hahn_2013, Gomez2014, Yu_2018}, we approximate $g^{\uparrow \downarrow}_m$ for Pt and Ta with YIG/Pt's: $g^{\uparrow \downarrow}_m \sim 10$~nm$^{-2}$ \cite{zhang2015role}.
\begin{figure}
	\includegraphics[width=3.25in]{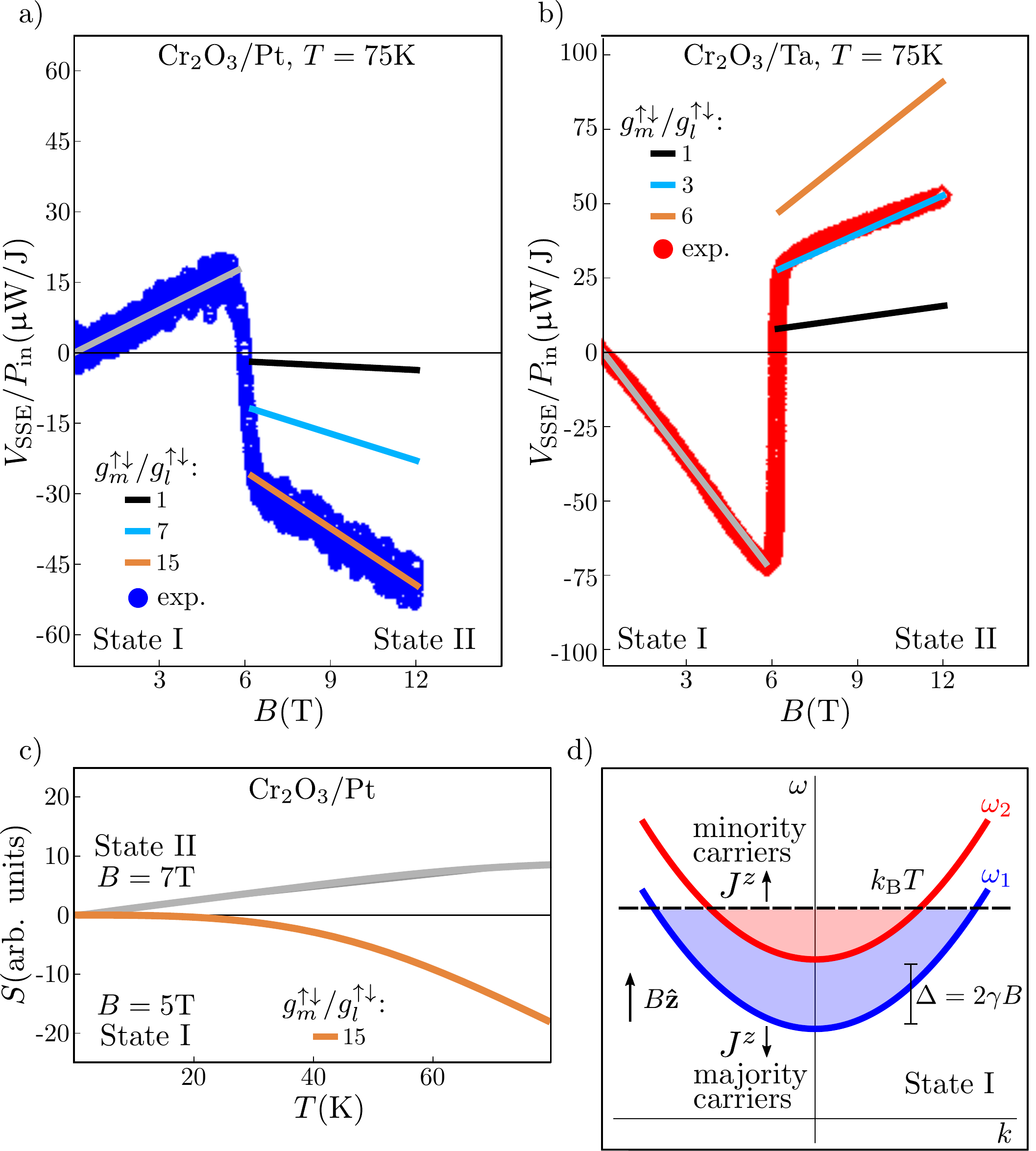}
	\caption{Theoretical spin Seebeck coefficients below, Eq.~\eqref{S_I}, and above, Eq.~\eqref{S_II}, spin flop for Cr$_2$O$_3$ are compared to experimental data from Li \textit{et al.} \cite{li2020}.  (a) and (b): The ratio $g^{\uparrow \downarrow}_m / g^{\uparrow \downarrow}_l$ is fit to the relative slopes across SF. c) $S(T)$ is plotted until $T = 80$~K; at higher temperatures, the long-wavelength theory loses quantitative accuracy. (d) Dispersions below SF are plotted. The majority spin carrier has magnetic moment along the field, which determines the polarization of the spin current.}
	\label{fig:fig_2}
\end{figure}
The comparison of the Seebeck coefficients \eqref{S_I}, \eqref{S_II} (which may be evaluated analytically \cite{s_I/II}) to the data \cite{li2020} is shown in Figs.~\ref{fig:fig_2}(a)-(b). We use the slope of experimental $V_{\mathrm{SSE}} / P_{\mathrm{in}}$ in I to determine $\kappa_{\mathrm{Pt}}^{*} \sim 10^9$~W/m$^2\cdot$K and $\kappa_{\mathrm{Ta}}^{*} \sim 10^{10}$~W/m$^2\cdot$K at $T = 75$~K, which are within 1-2 orders of magnitude of Stoner \textit{et al.} measurements \cite{Stoner_1992} of $\kappa$ in diamond$|$heavy-metal films. We also use an independent measurement of crystalline Cr$_2$O$_3$'s bulk thermal conductivity $\sigma$ \cite{yuan2018}, giving us an associated length scale $\sigma / \kappa_{\mathrm{Pt}}^{*} \approx 400$~nm and $\sigma / \kappa_{\mathrm{Ta}}^{*} \approx 60$~nm. Since the thin-film resistivities in our samples are about ten times larger than those in Refs. \cite{Vlaminck_2013, dutta_2017} for Pt, from which we use the values for $\lambda_{\mathrm{Pt}}^*$ and $g^{\uparrow \downarrow}_m$ which go into determining $\kappa_{\mathrm{Pt}}^{*}$, the latter can only be taken as giving us a rough order-of-magnitude guidance.

It should be safe to suppose that $\rho$, $\kappa^{*}$, and $g^{\uparrow \downarrow}$ are largely field independent, so that the field dependence in $V_{\mathrm{SSE}}/P_{\mathrm{in}}$ comes from $S$. The relative value of $S(B)$ across SF is determined theoretically up to the ratio $g^{\uparrow \downarrow}_m / g^{\uparrow \downarrow}_l$ \cite{gl_gm}, which is a property of the interfaces. Several values are chosen in plotting Fig. \ref{fig:fig_2}. The best fit is determined by comparing theoretical $v(T)$ \cite{s_I/II}, defined in Eq.~\ref{v_def}, to the data at $T = 75$~K. Note that $S|_{B = 0} = 0$, as expected on symmetry grounds. However, it is nontrivial that the $S_{\rm II}(B)$ dependence extrapolates to zero at zero field, both experimentally and in our theory. 

\begin{figure}
	\includegraphics[width=2.45in]{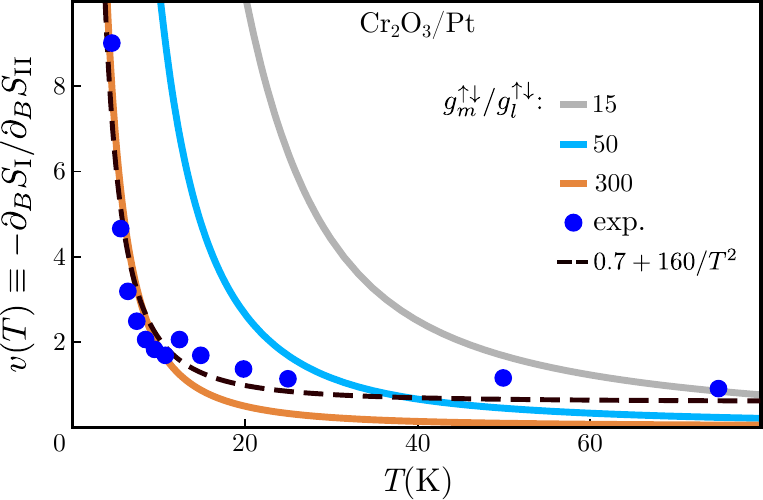}
	\caption{The ratio of the spin Seebeck coefficient field slopes $v(T)$. Experimental data is from the same device as in Fig.~\ref{fig:fig_2}(a) and is obtained from the slopes of linear-in-field fit lines, as discussed in the text. Theoretical curves are based on Eq.~\eqref{v_def}, evaluated here \cite{s_I/II}; plotted for various $g^{\uparrow \downarrow}_m / g^{\uparrow \downarrow}_l$. The dashed line shows an approximate fit to the data.}
	\label{fig:fig_3}
\end{figure}

The temperature dependence in the calculated spin Seebeck coefficient $S$ enter through the magnon occupation number in the fluctuation-dissipation relation \eqref{fdt}. The overall temperature dependence of the measured SSE is, furthermore, convoluted with thermal and charge conductivities. There are also slower temperature dependences in various parameters, such as $\chi(T)$ \cite{Foner_1963}, which can complicate a detailed analysis. By looking at the slope ratio $v(T)$, however, we can eliminate the common prefactor associated with the heat-to-spin-to-charge conversions [see Eq.~\eqref{VSSE}], if the signal is dominated by the interfacial thermal bias. The experimental $v(T)$ for a bulk Cr$_2$O$_3$/Pt sample is plotted in Fig.~\ref{fig:fig_3} along with theoretical curves. The experimental data points for $v(T)$ are obtained by fitting a linear-in-field line to $V_{\mathrm{SSE}}$ in states I and II and taking the ratio of the slopes; for the theoretical curves see \cite{s_I/II}. At low temperatures $T < 7$~K, the theoretical slopes start becoming nonlinear [so that $S_{\mathrm{I}}, S_{\mathrm{II}}$ must be evaluated numerically using Eqs.~\eqref{S_I},~\eqref{S_II}], with $S_{\mathrm{II}}(B)$ at large fields being the first portion of $S(B)$ to become nonlinear. Nonlinearities in $V_{\mathrm{SSE}}(B)$ are also observed experimentally above SF at $T = 5$~K \cite{li2020}. 

While we see qualitative agreement, it appears there are additional spin Seebeck contribution(s) not captured by our formalism.  The latter can stem from a bulk SSE in state I \cite{lebrun2018}, since thermal magnons polarized along the N\'eel order can diffuse over long distances \cite{Prakash_2018}. In particular, an additional linear in $T$ contribution to $S_\mathrm{I}$ would affect the estimate of $g^{\uparrow \downarrow}_m / g^{\uparrow \downarrow}_l$ from the low-$T$ data, while a cubic contribution would explain the constant offset in $v(T)$ at larger temperatures. There may also be additional contributions in I and II due to other types of dynamics associated with interfacial inhomogeneities and locally uncompensated moments. In order to fit the totality of experimental data with our interfacial SSE-based model, we would require different values of $g^{\uparrow \downarrow}_m / g^{\uparrow \downarrow}_l$ as a function of temperature. In particular, the data shown in Fig. \ref{fig:fig_2}a for $T = 75$~K (corresponding to the largest temperature data point in Fig. \ref{fig:fig_3}) is well reproduced by taking $g^{\uparrow \downarrow}_m / g^{\uparrow \downarrow}_l \approx 15$, while the low temperature dependence of the data follows $v(T) \approx 160/T^2$ corresponding to $g^{\uparrow \downarrow}_m / g^{\uparrow \downarrow}_l \approx 300$. Although the order-of-magnitude estimate for the mixing conductance ratio and the trend in $v(T)$ as a function of temperature are reasonably captured by our simple model, a more complete theory (accounting for the bulk spin transport as well as for disorder-induced mesoscopic effects at the interface) is needed for developing a detailed quantitative understanding.

\textit{Theoretical formalism.}|We calculate the spin currents in Eqs.~\eqref{g_def} by averaging over thermal fluctuations of the magnetic variables. The latter can be obtained from the symmetrized fluctuation-dissipation theorem:
\begin{equation}
\label{fdt}
\left\langle \delta \phi_i \delta \phi_j \right\rangle = \frac{i \hbar}{2} \int \frac{ d^3 k}{(2\pi)^3}\left[ \chi_{ji}^*(\mathbf{k},\omega) - \chi_{i j}(\mathbf{k},\omega)\right]N(\omega),
\end{equation}
where $\delta\phi_i$ stands for a Cartesian component of  $\boldsymbol{l}$ or $\boldsymbol{m}$ and $\chi_{ij}$ is the corresponding linear-response function. $N(\omega)\equiv n_{\rm BE}(\omega)+1/2$ accounts for thermal fluctuations associated with occupied modes, according to the Bose-Einstein distribution function $n_{\rm BE}$, with $1/2$ reflecting the zero-point motion \cite{landau1980statistical}. The dynamic susceptibility tensor is defined by $\delta \phi_i = \chi_{i j} \xi_j$, for the field $\xi_j$ thermodynamically conjugate to $\phi_j$. Our system is driven according to the energy density $E(B,t) = E(B) - \boldsymbol{m} \cdot \boldsymbol{h}(t) - \boldsymbol{l} \cdot \boldsymbol{g}(t)$, where $\boldsymbol{g}$ and $\boldsymbol{h}$ are conjugate to $\boldsymbol{l}$ and $\boldsymbol{m}$, respectively. The off-diagonal components of the N\'eel response $\chi^{(l)}_{i j}$ thus determine the N\'eel pumping as $\left\langle \boldsymbol{l} \times \partial \boldsymbol{l} / \partial t \right\rangle_k \to i \omega \epsilon^{ijk}\left\langle l_i l_j \right\rangle$ (in terms of the Levi-Civita tensor $\epsilon^{ijk}$, and upon the Fourier transform), and similarly for the magnetic response, $\chi^{(m)}_{i j}$.

The components contributing to spin currents in I are
\small
\begin{subequations}
	\begin{eqnarray}
	\label{chi_I}
	&&\chi_{x y}^{(l)} = -\frac{i}{2 s^2 \chi\omega_{0k}} \left( \frac{1}{\omega- \omega_{1k} + i\epsilon} - \frac{1}{\omega- \omega_{2k} + i\epsilon} \right),\\
	&&\chi_{x y}^{(m)} = \chi^2 K_1^2 \chi_{x y}^{(l)},
	\end{eqnarray}
\end{subequations} \normalsize
where $\omega_{0k} = \sqrt{(\gamma B_c)^2 + (ck)^2}$ and the dispersions are given in Eq.~\eqref{disp}. According to Eq.~\eqref{chi_I}, the fluctuations perpendicular to $\boldsymbol{l}_{0, \mathrm{I}} = \hat{\textbf{z}}$ at $\omega_{1k}$ and $\omega_{2k}$ produce opposite contributions to the spin currents. The magnetic fluctuations in I in, e.g. Cr$_2$O$_3$, are a factor $(\chi K_1)^2 \sim 10^{-7}$ smaller than the N\'eel fluctuations and will be neglected. In II, $\delta\boldsymbol{l}$ is linearly polarized in the $\omega_{3k}$ and $\omega_{4k}$ modes, so N\'eel fluctuations do not produce spin currents \cite{beyond_nonlinear_sigma}. $\delta\boldsymbol{m}$ is elliptically polarized in the $\omega_{4k}$ mode, with magnetic fluctuations producing a spin current according to
\begin{equation}
\chi_{x y}^{(m)} = i \gamma \chi B \left( \frac{1}{\omega- \omega_{4k} + i\epsilon} \right). \label{chi_II_mxmy}
\end{equation}

Without dissipation, the poles $\chi_{i j} \propto 1/(\omega - \omega_k + i\epsilon)$ at the resonance frequencies are shifted by positive infinitesimal $\epsilon$. With dissipation, we end up with Lorentzians centered at these poles, whose widths are determined by bulk Gilbert damping and the effective damping due to interfacial spin pumping \cite{tserkovnyak_2002, hoffman_2013}. When these resonance modes' quality factors are large, however, their spectral weight is sharp and may be simply integrated over. We will assume this is the case, allowing us to neglect dissipation and simply use the infinitesimal $\epsilon$.

\textit{Conclusion and outlook.}|Our theory specializes to SSE from spin currents produced by an interfacial thermal bias. The formalism may be extended to account for bulk thermal gradients, which produce nonequilibrium interfacial spin accumulation $\boldsymbol{\mu}$. However, determining  $\boldsymbol{\mu}$ requires complimenting the interfacial transport with coupled spin and heat transport in the bulk \cite{Prakash_2018}, which is beyond our present scope. The purely local SSE studied here should quantitatively model SSE for interfaces with large interfacial thermal resistances and weak interfacial spin coupling. In this regime, SSE would provide a noninvasive probe of the magnet's transverse components of $\chi_{i j}$, much like scanning tunneling microscopy is an interfacial probe of an electron density of states \cite{tersoff1983}.

We have discussed two classes of systems which produce different signs for SSE. The FM-like class involves spin excitations with magnetic moment opposite the order parameter, such as in FMs, uniaxial AFs above SF, and DMI AFs. Another class involves degenerate spin excitations, whose degeneracy is lifted by magnetic field. The majority carrier, which has magnetic moment along the magnetic field, can then dominates spin transport. In our low-temperature, long-wavelength theory we have shown that uniaxial AFs below SF belong to this class. However, when the bulk SSE contribution is significant, this reasoning alone may not determine the sign. Since the majority band reaches the edge of the BZ faster than the minority, it may suffer greater umklapp scattering at elevated temperatures, which would lower its conductivity. A full transport theory is then required to determine the SSE sign, as a function of temperature.

By comparing $v(T) \equiv - \partial_B S_{\mathrm{I}} / \partial_B S_{\mathrm{II}}$ in experiment to our theory as a function of $T$, we see some discrepancy. Our theory predicts $v \propto 1/T^2$, while the Cr$_2$O$_3$/Pt sample indicates $v(T) \approx 0.7+160/T^2$. The constant offset could stem from a bulk Seebeck contribution in I at higher $T$ whose coefficient goes as $T^3$. Above SF, bulk contributions to SSE can be expected to be reduced, since spin transport is then normal to the N\'eel order. $v(T)$ may also have contributions from paramagnetic impurities or other extrinsic surface modes, or be convoluted with temperature dependence in $g^{\uparrow \downarrow}_m / g^{\uparrow \downarrow}_l$. The magnitude of $g^{\uparrow \downarrow}_l$ and $g^{\uparrow \downarrow}_m$ can, furthermore, vary from one sample to another due to the amount of disorder in the interfacial exchange coupling \cite{Takei2014, troncoso2019}. While our theory well reproduces the temperature dependence at low $T$, a different value of $g^{\uparrow \downarrow}_m / g^{\uparrow \downarrow}_l$ is needed to consistently explain higher temperature data. Looking forward, a more complete theory is called for which includes SSE contributions from both the interface and the bulk, in addition to the dynamical effects of disorder at the interface.

The sensitivity of the SSE to the preparation and quality of the interface may complicate the analysis based on the measured $v(T)$ across the SF. We recall that Seki \textit{et al.} \cite{Seki_2015} did not observe a significant SSE in I at low temperatures in Cr$_2$O$_3$/Pt. Wu \textit{et al.} \cite{Wu_2016} observed SSE with nonlinear field dependence and ferromagnetic sign signature on both sides of SF in MnF$_2$/Pt. Ferromagnetic sign in I was also observed in an etched-interface Cr$_2$O$_3$/Pt sample by Li \textit{et al.} \cite{li2020}. Thus the origin of the measured sign of the signal in I, and, therefore, the physical mechanism of SSE are unclear for these cases. We also note that both Wu \textit{et al.} \cite{wu2015paramagnetic} in paramagnetic SSE in GGG/Pt and Li \textit{et al.} \cite{li2020} in Cr$_2$O$_3$/Pt at $T > T_N$ observed the ferromagnetic sign signature, suggesting perhaps the importance of the magnon umklapp scattering in the bulk.

The work was supported by the U.S. Department of Energy, Office of Basic Energy Sciences under Award No. DE-SC0012190.

\bibliography{sf}

\begin{thebibliography}{51}
\bibitem [{bey()}]{beyond_nonlinear_sigma}%
  \BibitemOpen
  \href@noop {} {}\bibinfo {note} {If we relax the nonlinear constraint $\delta
  \boldsymbol{l}^2 = 1$, we can allow for an additional term $ \boldsymbol{m}
  \times \delta E / \delta \boldsymbol{l}$ in the equation of motion for
  $\boldsymbol{l}$. When considering linear excitations about the same ground
  states, the only change then is that $\omega_{3k}$ develops small elliptical
  polarization in $\boldsymbol{\delta l}$. This produces a N\'eel spin current
  parallel to the field with similar magnitude to the $\omega_{4k}$ magnetic
  spin current. Since it pumps at $g^{\uparrow \downarrow}_l$ $\lesssim$
  $g^{\uparrow \downarrow}_m$, we discard it here.}\BibitemShut {Stop}%
\bibitem [{k2_()}]{k2_note}%
  \BibitemOpen
  \href@noop {} {}\bibinfo {note} {For example, in Cr$_2$O$_3$, the temperature
  associated with the zero-field magnon gap in I is $T = \hbar \gamma B_c / k_B
  \approx 8$~K, and the temperature associated with $K_2$ will be much less
  than this. So at all but low temperatures, the majority of magnons
  contributing to SSE will have frequencies which are unaffected by
  $K_2$.}\BibitemShut {Stop}%
\bibitem [{s_I()}]{s_I/II}%
  \BibitemOpen
  \href@noop {} {}\bibinfo {note} {For $k_BT\gg\hbar\gamma B_c$, we get
  \begin{equation*} S_{\mathrm{I}} \approx \frac{g^{\uparrow \downarrow}_l
  \gamma B k_B^2 T}{2 \pi^3 c^3 \chi s^2 } \int_0^\infty dx ~ x^2 e^x n_{\rm
  BE}^2(x) \propto g^{\uparrow \downarrow}_l B T,\end{equation*}
  \begin{equation*} S_{\mathrm{II}} \approx \frac{g^{\uparrow \downarrow}_m
  \gamma \chi B k_B^4 T^3}{4 \pi^3 c^3 \hbar^2 } \int_0^\infty dx ~ x^4 e^x
  n_{\rm BE}^2(x) \propto g^{\uparrow \downarrow}_m B T^3,\end{equation*} where
  $x$ is dimensionless and the integrals are convergent.}\BibitemShut {Stop}%
\end{thebibliography}
\end{document}